\documentclass[twocolumn,aps,pra,10pt]{revtex4-1}
\usepackage{bm}
\usepackage{amsfonts}
\usepackage{amssymb}
\usepackage{amsmath}
\usepackage{graphicx}
\newcommand{\sint}{\;\int\mspace{-26mu}\sum}
\newcommand{\sintt}{\;\int\mspace{-16mu}\Sigma}

\begin{document}

\title{Uncertainty relation for focal spots in light beams}
\author{Iwo Bialynicki-Birula}\email{birula@cft.edu.pl}
\affiliation{Center for Theoretical Physics, Polish Academy of Sciences\\
Aleja Lotnik\'ow 32/46, 02-668 Warsaw, Poland}
\author{Zofia Bialynicka-Birula}\affiliation{Institute of Physics, Polish Academy of Sciences\\
Aleja Lotnik\'ow 32/46, 02-668 Warsaw, Poland}

\begin{abstract}
Uncertainty relations for light pulses found in [Phys. Rev. A {\bf 86}, 022118 (2012)] were derived in the three-dimensional case which emphasized the localization in a volume. Here we derive the uncertainty relation for light beams in the two-dimensional plane perpendicular to the direction of the beam propagation which is more interesting for realistic beams. This uncertainty relation connects the area of the focal spot with the spectrum of transverse photon momenta. The shape of the beam that saturates the uncertainty relation is very close to a Gaussian. The directions of the electric and magnetic field vectors are those of the circularly polarized plane wave. Our uncertainty relation for the focal spot is quite general but we were able to determine the value of the lower bound only for beams made of many photons.
\end{abstract}

\maketitle

\section{Introduction}

Quantum-mechanical properties of photons impose a relationship between the spatial and the spectral properties of light beams. It may be given the form of uncertainty relations derived in \cite{bb1,bb2}. The uncertainty relations for light beams follow from the fundamental properties of relativistic quantum systems and they cannot be overcome by technical improvements introduced in the process of the beam formation.

In this paper we analyze the restrictions imposed on the area of the focal spot (beam waist) by the Heisenberg-type uncertainty relation that involves the variance of the transverse wave vector. We fully confirm our intuitive feeling that the better the beam is focused the less certain is the transverse beam momentum. To make the calculations feasible we assume that the beam is intense, i. e. the average number of photons $N$ is very large. This enables us to simplify the calculations by keeping only the leading terms in the expansion in $1/N$.

There is some overlap between this work and our previous papers. However, there we mostly dealt with the uncertainty relations for single photons. The treatment of light beams in the last section of Ref.~\cite{bb2} was rather sketchy and it dealt only with light pulses and not with light beams since we studied exclusively the three-dimensional case. Here we derive the two-dimensional uncertainty relation which is more appropriate for light beams.

There are many different ways to express the physical uncertainty relations in terms of mathematical inequalities. For example, in the nonrelativistic quantum mechanics we have the standard Heisenberg relations but also the entropic uncertainty relations. In the case of beams of light we have an almost universal measure of the directional spread of wave vectors but many measures of the spatial spread. Siegman \cite{sieg} listed seven possible definitions of the beam width ``that have been suggested or used for optical beams in the past.'' However, only one or two might be useful for the derivation of uncertainty relations. In this context we must mention the uncertainty relations derived earlier by M. A. Alonso and his collaborators \cite{al1,al2,al3,al4} that are based on a completely different measure of the spatial spread. We shall compare their results with ours in the Conclusions.

To introduce the main concepts, let us consider a well collimated beam of light centered on the $z$ axis. Following \cite{bb2}, we shall measure the beam width with the use of the center of energy operator ${\hat{\bm R}}$ introduced almost eighty years ago by Born and Infeld \cite{bi}. Even though the components of this operator do not commute, it is still the best available substitute for the nonexisting genuine photon position operator. The size of the beam waist will be measured by the variance of the center of energy operator $\hat{\bm R}$ in the transverse plane,
\begin{align}\label{varr}
\Delta{\bm R}_\perp^2=\langle {\bm R}_\perp^2\rangle-\langle{\bm R}_\perp\rangle^2
\end{align}
and the spread of the transverse momentum by the variance of the transverse momentum operator:
\begin{align}\label{varp}
\Delta{\bm P}_\perp^2=\langle {\bm P}_\perp^2\rangle-\langle{\bm P}_\perp\rangle^2.
\end{align}
We prove that the following uncertainty relation holds:
\begin{align}\label{ur}
\sqrt{\Delta{\bm R}_\perp^2}\sqrt{\Delta{\bm P}_\perp^2}\ge\gamma_\perp\hbar.
\end{align}
The dimensionless parameter $\gamma_\perp$ depends on the mean value $\langle k_z\rangle$ of the component of the wave vector in the beam direction. In the limit, when $\langle k_z\rangle$ tends to infinity, $\gamma_\perp$ tends to 1. In the process of proving this relation we find the wave fields that saturate the inequality.

The use of the center of energy operator to characterize the spatial extension of a {\em single photon} may seem awkward. However, its use for intense light beams should not raise doubts because it is the energy distribution in a beam that gives the best characteristic of the beam shape.

\section{Center of the energy operator}

In classical electrodynamics we define the center of the energy $\bm R$ of the electromagnetic field as the following ratio:
\begin{align}\label{cecl}
{\bm R}=\frac{\int d^3r\,{\bm r}{\mathcal E}(\bm r)}{\int d^3r\,{\mathcal E}(\bm r)},
\end{align}
where ${\mathcal E}(\bm r)$ is the energy density. In quantum theory the energy density becomes an operator and to obtain a Hermitian operator $\hat{\bm R}$ we have to symmetrize the product of two noncommuting operators,
\begin{align}\label{ceq}
\hat{\bm R}=\frac{1}{2\hat H}\hat{\bm N}+\hat{\bm N}\frac{1}{2\hat H},
\end{align}
where we identified the total energy operator with the Hamiltonian,
\begin{align}\label{ham}
\hat{H}={\int d^3r\,\hat{\mathcal E}(\bm r)},
\end{align}
and we denoted the first moment of the energy by $\hat{\bm N}$,
\begin{align}\label{nop}
\hat{\bm N}={\int d^3r\,{\bm r}\hat{\mathcal E}(\bm r)}.
\end{align}
Since the momentum operator of the electromagnetic field acts as the generator of translations,
\begin{align}\label{tr}
\left[\hat{\bm P},\hat{\mathcal E}(\bm r)\right]=i\hbar\bm{\nabla}\hat{\mathcal E}(\bm r),
\end{align}
the operators ${\hat{\bm R}}$ and ${\hat{\bm P}}$ obey the canonical commutation relations for position and momentum,
\begin{align}\label{can}
[{\hat{R}_i},{\hat{P}_j}]=i\hbar\delta_{ij}.
\end{align}
There is, however, one disconcerting property of the center of the energy operator: its components do not commute. This follows directly from the commutator of the energy densities,
\begin{align}\label{fund}
[\hat{\mathcal E}(\bm r),\hat{\mathcal E}(\bm r')]=-i\hbar c^2\left[\hat{\mathcal P}^k(\bm r)-\hat{\mathcal P}^k(\bm r')\right]\partial_k\delta(\bm r-\bm r'),
\end{align}
where $\hat{\bm{\mathcal P}}(\bm r)$ is the density of momentum. This relation was called by Schwinger \cite{js} ``the most fundamental equation of relativistic quantum field theory'' since it must hold for every relativistic system. With its use we obtain first
\begin{align}\label{nnm}
{\hat{\bm N}}\times{\hat{\bm N}}=-i\hbar c^2\hat{\bm M},
\end{align}
and then \cite{bi},
\begin{align}\label{rcom}
{\hat{\bm R}}\times{\hat{\bm R}}=-i\hbar c^2\frac{1}{\hat H}\hat{\bm S}\frac{1}{\hat H},
\end{align}
where the spin operator $\hat{\bm S}$ is defined as the difference between the total angular momentum $\hat{\bm M}$,
\begin{align}\label{am}
\hat{\bm M}={\int d^3r\,{\bm r}\times\hat{\bm{\mathcal P}}(\bm r)},
\end{align}
and the orbital angular momentum $\hat{\bm R}\times\hat{\bm P}$. In the case of photons (or the electromagnetic field) the role of spin is played by helicity.

It is worthwhile to note that the complete list of commutators for the operators $\hat{H}, \hat{\bm P}, \hat{\bm M}$, and $\hat{\bm N}$ coincides with the commutator algebra for the Poincar\'e group with $\hat{\bm N}$ being the generator of Lorentz transformations. Thus, noncommutativity of the components of $\hat{\bm R}$ is a direct consequence of the noncommutativity of Lorentz transformations. This noncommutativity  does not preclude the use of $\hat{\bm R}$ to characterize the spatial extension of the system. In a relativistic system there is no other quantity that can better serve as a replacement for the nonrelativistic position.

\section{Light beams as coherent states of the electromagnetic field}

To describe light beams we choose the coherent states: the most classical states of the electromagnetic field. Coherent states $|\text{coh}\rangle$ are constructed \cite{rg,*tg,*sr} by acting on the vacuum state vector with the unitary Glauber displacement operator $D$,

\begin{align}\label{coh}
|\text{coh}\rangle=D|0\rangle,
\end{align}
\begin{align}
D=\exp\left(\!\sqrt{N}\sint\!\left[f_\lambda(\bm k)a_\lambda^\dagger(\bm k)-f_\lambda^*(\bm k)a_\lambda(\bm k)\right]\right),
\end{align}
where the symbol $\sintt$ stands for both the integration over all wave vectors $\int\!d^3k/k$ and the summation over two circular polarizations $\lambda=\pm 1$. The factor $1/k$ transforms the volume element $d^3k$ into the relativistically invariant measure $d^3k/k$ on the light-cone \cite{pp}.

The mode functions $f_\lambda(\bm k)$ fully describe the shape of the light beam. We assume the invariant normalization of the mode function,
\begin{align}\label{norm}
\sint\!|f_\lambda(\bm k)|^2=1.
\end{align}
With this normalization of $f_\lambda(\bm k)$ the number $N$ appearing in (\ref{coh}) is equal to the expectation value of the photon number operator ${\hat{N}}$,
\begin{align}\label{nn}
{\hat{N}}=\sint a_\lambda^\dagger(\bm k) a_\lambda(\bm k),
\end{align}
in the coherent state, $N=\langle{\hat{N}}\rangle$. Our relativistic convention implies that the commutation relations between the annihilation and creation operators contain an extra factor of $k$,
\begin{align}\label{cr1}
\left[a_\lambda(\bm k),a_{\lambda'}^\dagger(\bm k')\right]=\delta_{\lambda\lambda'}k\,\delta^{(3)}(\bm k-\bm k').
\end{align}
The displacement operator shifts the creation and annihilation operators by the following c-number terms:
\begin{subequations}\label{dis}
\begin{align}
D^\dagger a^\dagger_\lambda(\bm k)D
=a^\dagger_\lambda(\bm k)+\sqrt{N}f_\lambda^*(\bm k),\\
D^\dagger a_\lambda(\bm k)D
=a_\lambda(\bm k)+\sqrt{N}f_\lambda(\bm k).
\end{align}
\end{subequations}
The explicit appearance of $N$ in these formulas simplifies the extraction of the leading terms in the uncertainty relations for intense beams.

\section{Measures of uncertainty in position and momentum}

In order to evaluate the variances (\ref{varr}) and (\ref{varp}) we must express ${\hat{H}}$, ${\hat{\bm P}_\perp}$, and ${\hat{\bm N}_\perp}$ in terms of creation and annihilation operators \cite{bb,*bb3,*bb4,bb5},
\begin{subequations}
\begin{align}\label{exp}
{\hat{H}}&=\sint\hbar\omega  a_\lambda^\dagger(\bm k)a_\lambda(\bm k),\\
{\hat{\bm P}_\perp}&=\sint\hbar{\bm k}_\perp  a_\lambda^\dagger(\bm k) a_\lambda(\bm k),\\
{\hat{\bm N}_\perp}&=\sint\hbar\omega a_\lambda^\dagger(\bm k)i{\bm D}_{\lambda\perp}a_\lambda(\bm k),
\end{align}
\end{subequations}
where ${\bm D}_\lambda$ is the covariant derivative on the light cone.
\begin{align}\label{d}
{\bm D}_\lambda={\bm\nabla}_{\bm k}-i\lambda{\bm\alpha}(\bm k),
\end{align}
and ${\bm\alpha}(\bm k)$ is the analog of the affine connection,
\begin{align}\label{ac}
{\bm\alpha}(\bm k)=\frac{k_z(-k_y,k_x,0)}{k(k_x^2+k_y^2)}.
\end{align}
The expression for $\Delta{\bm P}_\perp^2$ obtained with the use of (\ref{dis}) contains only one term,
\begin{align}\label{delp}
\Delta{\bm P}_\perp^2=N\sint \hbar^2{\bm k}_\perp^2 |f_\lambda(\bm k)|^2,
\end{align}
because the terms proportional to $N^2$ cancel between the two parts of the variance present in (\ref{varp}).

In order to calculate $\Delta{\bm R}_\perp^2$ we need the following formulas for $D^\dagger{\hat{\bm N}_\perp}D$ and $D^\dagger(1/{\hat{H}})D$:
\begin{widetext}
\begin{subequations}
\begin{align}\label{corr}
D^\dagger {\hat{\bm N}_\perp}D&=N\left[\sint\hbar\omega f_\lambda^*(\bm k) i{\bm D}_{\lambda\perp} f_\lambda(\bm k)
+\frac{1}{\sqrt{N}}\sint\hbar\omega\,\left(a^\dagger_\lambda(\bm k)i{\bm D}_{\lambda\perp} f_\lambda(\bm k)+f^*_\lambda(\bm k)i{\bm D}_{\lambda\perp} a_\lambda(\bm k)\right)+{\mathcal O}\left(\frac{1}{N}\right)\right],\\
D^\dagger\frac{\hbar c}{\hat{H}}D&=\frac{1}{N\mathcal H}\left[1-\frac{1}{\mathcal H\sqrt{N}}
\sint\,k\left(a^\dagger_\lambda(\bm k)f_\lambda(\bm k)
+f^*_\lambda(\bm k) a_\lambda(\bm k)\right)+{\mathcal O}\left(\frac{1}{N}\right)\right],
\end{align}
\end{subequations}
\end{widetext}
where $\mathcal H$ is the average value of $k$ per one photon in the coherent state,
\begin{align}\label{clen}
\mathcal H=\sint k\,|f_\lambda(\bm k)|^2.
\end{align}
The evaluation of $\Delta{\bm R}_\perp^2$ in the lowest order is now simple but tedious. Again the leading terms cancel between the two terms in the variance (\ref{varr}) resulting in:
\begin{align}\label{finr}
&\Delta{\bm R}_\perp^2=\frac{1}{N{\mathcal H}^2}\sint k^2\nonumber\\
&\times\left[(i{\bm D}_{\lambda\perp} -{\bm{\mathcal R}_\perp})f_\lambda(\bm k)\right]^*\!\cdot\!(i{\bm D}_{\lambda\perp} -{\bm{\mathcal R}_\perp})f_\lambda(\bm k),
\end{align}
where
\begin{align}\label{r}
{\mathcal R}_\perp=\frac{1}{\mathcal H}\sint k f_\lambda^*(\bm k) i{\bm D}_{\lambda\perp} f_\lambda(\bm k).
\end{align}
We can eliminate ${\bm{\mathcal R}_\perp}$ by changing the phase of $f_\lambda(\bm k)$.
\begin{align}\label{phase}
f_\lambda(\bm k)\to e^{-i{\bm k}\cdot{\bm{\mathcal R}_\perp}} f_\lambda(\bm k).
\end{align}
This change of the phase in momentum space is equivalent to the translation by ${\bm{\mathcal R}_\perp}$ in the coordinate space. Since $\langle{\bm R}_\perp\rangle={\bm{\mathcal R}_\perp}+{\mathcal O}\left(\frac{1}{N}\right)$, after the translation the origin of the coordinate system coincides with the center of energy. Alternatively, we could have chosen the coordinate system in this way from the very beginning and then $\langle{\bm R}_\perp\rangle$ would be equal to 0.

Finally, the left hand side of the uncertainty relation takes on the form:
\begin{align}\label{finur}
\gamma_\perp^2&=\frac{\Delta{\bm R}_\perp^2\Delta{\bm P}_\perp^2}{\hbar^2}\nonumber\\
&=\frac{\sintt\,k^2
|{\bm D}_{\lambda\perp} f_\lambda(\bm k)|^2\,\sintt\,{\bm k}_\perp^2 |f_\lambda(\bm k)|^2}{\left[{\sintt\,k\,|f_\lambda(\bm k)|^2}\right]^2}.
\end{align}
We replaced momenta by the corresponding wave vectors since for intense light beams the Planck constant does not play a significant role. In the next section we apply the variational procedure to find the minimum of $\gamma_\perp$.

\section{The variational equation}

Up to this point we followed the general procedure developed previously for the three-dimensional case. However, the formula (\ref{finur}) for $\gamma_\perp^2$ differs from the formula (66) for $\gamma^2$ in \cite{bb2} so that the variational equation for $f_\lambda(\bm k)$ will now be different. Variation of $\gamma_\perp^2$ with respect to $f_\lambda^*(\bm k)$ leads to an equation that allows for the separation of variables in polar coordinates. After the substitution $f_\lambda(\bm k)=\exp(iM\phi)f_\lambda(k_\perp,k_z)$ we obtain the following equation for $f_\lambda(k_\perp,k_z)$:
\begin{align}\label{var}
{\mathcal D}f_\lambda(\kappa_\perp,\kappa_z)=0,
\end{align}
where ${\mathcal D}$ is the following differential operator:
\begin{widetext}
\begin{align}\label{dvar}
{\mathcal D}=-\partial_{\kappa_\perp}^2-\left(\frac{1}{\kappa_\perp}
+\frac{\kappa_\perp}{\kappa_\perp^2+\kappa_z^2}\right)\partial_{\kappa_\perp}
+\frac{1}{\kappa_\perp^2}\!\left(\frac{\kappa_z}{\sqrt{\kappa_\perp^2+\kappa_z^2}}
-\lambda M\right)^2
+\gamma_\perp^2\left(\frac{\kappa_\perp^2}{\kappa_\perp^2+\kappa_z^2}
-\frac{2}{\sqrt{\kappa_\perp^2+\kappa_z^2}}\right).
\end{align}
\end{widetext}
We introduced here the dimensionless variables,
\begin{align}\label{dim}
\kappa_\perp=\frac{k_\perp\mathcal H}{\Delta{\bm k}_\perp^2},\quad
\kappa_z=\frac{k_z\mathcal H}{\Delta{\bm k}_\perp^2}.
\end{align}
It is sufficient to consider one polarization at a time, say $\lambda=1$, because the two functions $f_\pm$ satisfy separate equations. The variational equation (\ref{var}) contains $\kappa_z$ only as a parameter. This equation is linear, therefore, every solution can be multiplied by an arbitrary function of $\kappa_z$, say $A(\kappa_z)$. Such a multiplicative factor is needed, in particular, to secure normalizability.

\section{Slim beams}

We shall characterize beam shapes with the use of the following dimensional parameters:
\begin{itemize}
\item{The standard deviation $d_\perp=\sqrt{\Delta{\bm k}_\perp^2}$ that measures the spread of the transverse components of the wave vector (the beam width in momentum space).}
\item{The average value $\langle\kappa_z\rangle$ of the longitudinal component of the wave vector.}
\item{The standard deviation $d_z=\sqrt{\Delta k_z^2}$ that measures the width of the distribution of $k_z$, as determined mainly by the amplitude $A(\kappa_z)$.}
\end{itemize}
Two dimensionless ratios: $s=d_\perp/\langle\kappa_z\rangle$ and $w=d_z/\langle\kappa_z\rangle$ characterize the properties of the beam in a scale-independent way. The parameter $s$ determines how slim the beam is. The parameter $w$ determines how well the average value $\langle\kappa_z\rangle$ represents all the values of $k_z$.

In most cases one is interested in (almost) monochromatic beams. Therefore, both $s$ and $w$ must be very small as compared to 1. We shall call the beams that satisfy the conditions $s\ll 1$ and $w\ll 1$ the {\em slim beams}. Note that for slim beams $\langle\kappa_z\rangle$ behaves as $1/s^2$ because $\mathcal H$ in (\ref{dim}) is roughly equal to $\langle k_z\rangle$. In what follows we often consider the limit when $w\to 0$, i.e. we neglect the width of the distribution $d_z$. We may then identify $\kappa_z$ in the variational equation with its average value $\langle\kappa_z\rangle$, dropping the average symbol $\langle\rangle$ altogether. Sometimes, however, for the sake of clarity we will keep the distinction.

\begin{figure}
\centering
\includegraphics[scale=0.8]{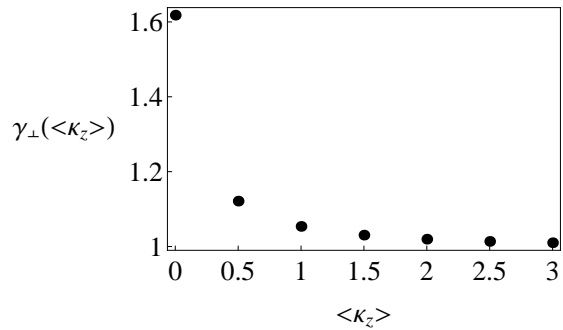}
\caption{Numerical data points for $\gamma_\perp$ as a function of the dimensionless longitudinal component of the wave vector $\langle \kappa_z\rangle$. With the increasing $\langle \kappa_z\rangle$ the value of $\gamma_\perp$ drops down very fast to 1.}\label{fig1}
\end{figure}
\begin{figure}
\centering
\includegraphics[scale=0.8]{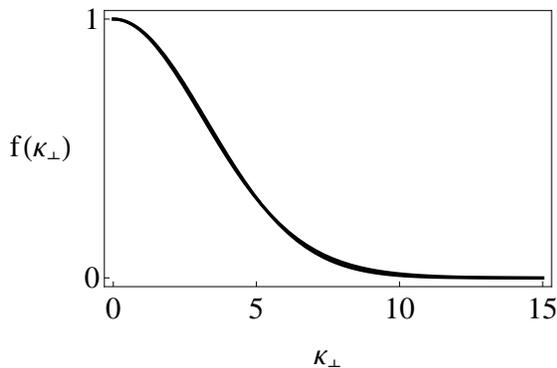}
\caption{Plots of the Gaussian function $\exp(-\kappa_\perp^2/2\langle\kappa_z\rangle)$ and the numerical solution $f(\kappa_\perp)$ of the variational equation corresponding to the value of the dimensionless parameter $\langle\kappa_z\rangle$=100. The difference between these two curves are hidden in the line width.}\label{fig2}
\end{figure}
It is worth noting that the $1/\kappa_\perp^2$ term in the variational equation (\ref{var}) corresponds to a centrifugal repulsion that pushes the function $f(\kappa_\perp^2,\kappa_z)$ away from the region of small $1/\kappa_\perp^2$. The presence of this term clearly increases the variance $\Delta k_\perp^2$, resulting in the increase of the lower bound in the uncertainty relation. To decrease the role of this repulsion we should take $M=\lambda=1$ or $M=\lambda=-1$ and a large value of $\langle k_z\rangle$. The other choice $M=0$ and $\kappa_z=0$ also eliminates the centrifugal term but the vanishing of $\kappa_z$ means that we no longer deal with a beam.

In Fig.~\ref{fig1} we show the lower bound $\gamma_\perp$ as a function of the average value of longitudinal component of the wave vector. It is seen that the value $\gamma_\perp=1$ (characteristic for nonrelativistic quantum mechanics in two dimensions) is obtained when $\kappa_z\to\infty$. For $\kappa_z=0$ the variational equation (\ref{var}) can be solved analytically giving $\gamma_\perp(0)=(1+\sqrt{5})/2\approx 1.618$. In Fig.~\ref{fig2} we show for $\kappa_z=100$ the function $f(\kappa_\perp,\kappa_z)$ that solves the variational equation (with the normalization such that $f(0,\kappa_z)=1$) compared with the Gaussian function $g(\kappa_\perp,\kappa_z)$,
\begin{align}\label{gauss}
g(\kappa_\perp,\kappa_z)=\exp\left(-\frac{\kappa_\perp^2}{2\kappa_z}\right)
=\exp\left(-\frac{k_\perp^2}{2\Delta{\bm k}_\perp^2}\frac{\mathcal H}{k_z}\right).
\end{align}
These two functions are practically identical. The Gaussian function gives rise to $\gamma_\perp=1.00007$. This value is only slightly higher than $\gamma_\perp=1.000024$ obtained for the numerical solution.

In the Appendix we give the proof that indeed the exact solutions tends very fast to the Gaussian function with the increase of $\kappa_z$. Therefore, we may freely replace the function that saturates the uncertainty relation by the corresponding Gaussian. In the next Section we describe the properties of the electromagnetic field corresponding to this choice.

\section{Electromagnetic field that saturates the uncertainty relation}

We shall calculate now in coordinate space the electromagnetic field that minimizes the left hand side of the uncertainty relation. The quantized electromagnetic field can be expressed in terms of the creation and annihilation operators through the Riemann-Silberstein vector \cite{bb5},
\begin{align}\label{frep0}
{\hat{\bm F}}(\bm r,t)&=\frac{{\hat{\bm D}}}{\sqrt{2\epsilon}}+i\frac{{\hat{\bm B}}}{\sqrt{2\mu}}
=\int\!\frac{d^3k}{(2\pi)^{3/2}}{\bm e}(\bm k)\nonumber\\
&\times\left[a_+(\bm k)e^{i\bm k\cdot\bm r-i\omega t}+a_-^\dagger(\bm k)e^{-i\bm k\cdot\bm r+i\omega t}\right],
\end{align}
where
\begin{align}\label{ee}
{\bm e}({\bm k}) = \frac{1}{\sqrt{2}\,k\sqrt{k_x^2+k_y^2}}\!\left[
\begin{array}{c}
-k_x k_z+i k k_y\\
-k_y k_z-i k k_x\\
k_x^2+k_y^2
\end{array}
\right].
\end{align}
The expectation value of the field operator in the coherent state (\ref{coh}) gives the connection between the field and the functions $f_\lambda(\bm k)$,
\begin{align}\label{frep}
&{\bm F}(\bm r,t)=\frac{\bm D}{\sqrt{2\epsilon}}+i\frac{\bm B}{\sqrt{2\mu}}=\sqrt{N}\int\!\frac{d^3k}{(2\pi)^{3/2}}{\bm e}(\bm k)\nonumber\\
&\times\left[f_+(\bm k)e^{i\bm k\cdot\bm r-i\omega t}+f_-^*(\bm k)e^{-i\bm k\cdot\bm r+i\omega t}\right],
\end{align}
We will take only the part with $\lambda=1$. Since the Gaussian function serves as a very good approximation, as seen in Fig.~\ref{fig2}, we shall use this analytic representation to determine the field in the coherent state that minimizes the uncertainty relation. We will again consider slim beams not only to make the Gaussian approximation applicable but also to make the integration in (\ref{frep}) feasible. First, we use the smallness of $w$ to perform the integration over $k_z$ and substitute the average value $\langle k_z\rangle$ in place of the integration variable $k_z$. In the remaining two-dimensional integral we choose the polar coordinates. Since we have chosen $\lambda=1$ we must also take $m=1$ and we are left with the function $e^{i\phi}e^{-\alpha k_\perp^2/2}$. The parameter $\alpha$ (as given by the formula (\ref{gauss})) for slim beams becomes $\alpha=1/\Delta{\bm k}_\perp^2$. Thus, we end up with the following integral over the polar variables $\phi$ and $k_\perp$:
\begin{align}\label{rep}
&{\bm F}(\rho,\varphi,z,t)\nonumber\\
&=\int_0^\infty\!\!dk_\perp k_\perp\int_0^{2\pi}\!\!\!d\phi\,
\frac{1}{k}\left[\begin{array}{c}
k_z\cos\phi-ik\sin\phi\\
k_z\sin\phi+ik\cos\phi\\
-k_\perp
\end{array}\right]\nonumber\\
&\qquad\times e^{i\phi}e^{-\alpha k_\perp^2/2}e^{i\rho k_\perp\cos(\phi-\varphi)}e^{i\langle k_z\rangle z}e^{-ikct},
\end{align}
where $\phi-\varphi$ is the angle between the vectors $\bm k_\perp$ and $\bm r_\perp$, $k=\sqrt{k_\perp^2+\langle k_z\rangle^2}$. We dropped an irrelevant overall field amplitude. After the shift of the integration variable $\phi\to\phi+\varphi$ the integration over $\phi$ can be performed (Eq.~8.411 of \cite{gr}), leading to the following integral over $k_\perp$ involving Bessel functions:
\begin{align}\label{rep1}
&{\bm F}(\rho,\varphi,z,t)\nonumber\\
&=\!\int_0^\infty\!\!dk_\perp k_\perp\frac{1}{k}\left[\!\begin{array}{c}
k_+J_0(k_\perp\rho)+k_-e^{2i\varphi}J_2(k_\perp\rho)\\
ik_+J_0(k_\perp\rho)-ik_-e^{2i\varphi}J_2(k_\perp\rho)\\
-ik_\perp e^{i\varphi}J_1(k_\perp\rho)
\end{array}\!\right]\nonumber\\
&\qquad\times e^{i\langle k_z\rangle z}e^{-ikct}e^{-\alpha k_\perp^2/2},
\end{align}
where $k_{\pm}=(k\pm\langle k_z\rangle)/2$. Note that the integrand in this formula is nothing else but the Bessel beam for $M=1$, cf., for example \cite{bb6}. Therefore the field is the superposition of Bessel beams with different $k_\perp$ weighted by the Gaussian function.

The properties of slim beams enable us to replace $k$ by $\langle k_z\rangle$ and to disregard the terms proportional to $k_-\approx\langle k_z\rangle s^2/2$ as compared to the terms proportional to $k_+\approx 2\langle k_z\rangle$. The integration over $k_\perp$ can now be done (Eq.~6.631 of \cite{gr}) and the final result is (up to normalization):
\begin{align}\label{rep2}
{\bm F}(\rho,\varphi,z,t)
=\left[\begin{array}{c}
1\\
i\\
-i\frac{\rho e^{i\varphi}\Delta{\bm k}_\perp^2}{\langle k_z\rangle}
\end{array}\right]
e^{-\rho^2\Delta{\bm k}_\perp^2/2}
e^{i\langle k_z\rangle(z-ct)}.
\end{align}
Thus, within our approximation, this beam is non-diffractive. The energy distribution in the transverse plane is axially symmetric and has the Gaussian form,
\begin{align}\label{en}
{\mathcal E}(\rho)={\bm F}^*\!\cdot\!{\bm F}=\left[2+\left(\frac{\rho\Delta{\bm k}_\perp^2}{\langle k_z\rangle}\right)^2\right]e^{-\rho^2\Delta{\bm k}_\perp^2}.
\end{align}
\begin{figure}
\centering
\includegraphics[scale=0.35]{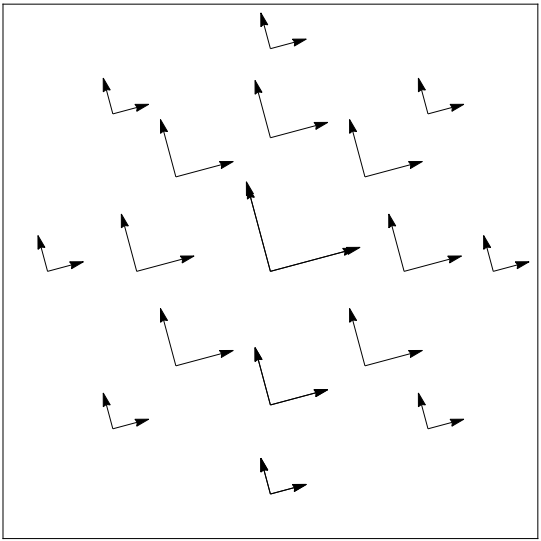}
\caption{Pairs of electric and magnetic field vectors. Their direction is the same in the whole transverse plane and their orientation is controlled by the phase factor $e^{i\langle k_z\rangle(z-ct)}$. The field strength decreases as we move away from the center.}\label{fig3}
\end{figure}
On the beam axis the electric field and the magnetic field are those of the circularly polarized plane wave. As we move away from the center the longitudinal component appears (it is needed to make the field divergenceless). However, the longitudinal component is practically negligible as compared with the transverse components because its maximal value $\sqrt{\Delta{\bm k}_\perp^2}/2\langle k_z\rangle$  is much smaller than 1. The electric and magnetic field vectors at each point form an orthogonal pair that has the same orientation in the whole transverse plane, as shown in Fig.~{\ref{fig3}. The field configuration (\ref{rep2}) is markedly different from the so called radially polarized beams studied in \cite{qdegl,*yb,*dql,*yy,*op}. Therefore, radially polarized beams do not have the smallest possible focal area.

\section{Conclusions}

We derived the uncertainty relation that imposes the lower limit on the focusing area. The size of this area is determined by the distribution of energy (beam intensity). The lowest value of the focal area is determined by the spectral properties of the beam; it is inversely proportional to the variance of the transverse components of the wave vectors present in the beam. In the limit of very slim beams the uncertainty relation becomes the same as the two-dimensional Heisenberg uncertainty relation in nonrelativistic quantum mechanics. We also determined the shape of the beam that saturates the uncertainty relation, i. e. the beam that is maximally focused given the value of the variance of transverse wave vectors. The energy density of this beam has a Gaussian shape while the directions of the field vectors (except for a small longitudinal component) are the same as in the circularly polarized plane wave (Fig.\ref{fig3}).

The main difference between our results and those described in Refs.\cite{al1,al2,al3,al4} is the use of a different measure of the spatial spread. The spread is measured in these works in a novel way (it is not on the Siegman's list), namely, by using the square of the angular part of angular momentum. This measure can only be applied to monochromatic beams of light (cf. Eq.(8) of Ref.\cite{al4}). In contrast, we use the measure of the beam width which is {\em not restricted} to the ideal case of monochromatic beams. Since it is based on the variance of the position operator (defined as the center of energy) our measure is as close as possible to the standard quantum mechanical definition. It seems to be also directly related to the observed intensity distribution in the beam. The second difference is the use of the covariant derivative (\ref{d}) on the light cone. The use of plain derivatives leads to the violation of divergence condition as seen in the formula (12) of Ref.\cite{al4}. Also, our uncertainty relation seems to be more fundamental since it is rooted in the relativistic properties of the electromagnetic field but that does not necessarily mean that it would be more useful in practical applications.

\section*{Acknowledgments}
This research was partly supported by the grant from the Polish Ministry of Science and Higher Education for the years 2013--2016.

\appendix

\section{}

As a natural measure of the difference between the Gaussian function and the exact solution of the variational equation we take the ratio $I(k_z)$ of the integral of the square of the function ${\mathcal D} g(\kappa_\perp,\kappa_z)$ divided by the norm of the Gaussian function,
\begin{align}\label{dep}
I(\kappa_z)=\frac{\int_0^\infty\!d\kappa_\perp[{\mathcal D} g(\kappa_\perp,\kappa_z)]^2}{\int_0^\infty\!d\kappa_\perp [g(\kappa_\perp,\kappa_z)]^2}.
\end{align}
\begin{figure}
\centering
\includegraphics[scale=0.8]{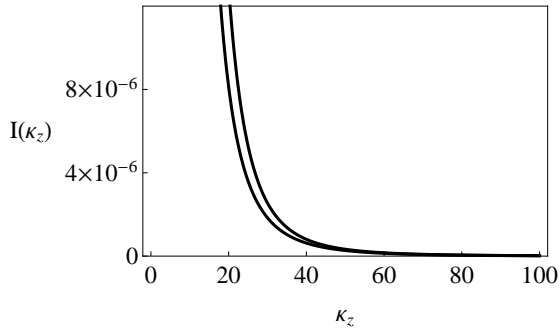}
\caption{This figure illustrates two facts: a) the Gaussian function approaches very fast the exact solution of the variational equation (lower curve) and b) the dependence on the dimensionless parameter $\kappa_z$ is very well approximated by the asymptotic formula (\ref{asym}) (upper curve).}\label{fig4}
\end{figure}
Obviously, when this integral is equal to 0, the function $g(\kappa_\perp,\kappa_z)$ obeys the variational equation (\ref{var}). This integral can be evaluated in a closed form \cite{math},
\begin{align}\label{dep1}
I(\kappa_z)&=c_0+c_1\,e^{\kappa_z}K_0(\kappa_z/2)+c_2\,e^{\kappa_z/2}K_1(\kappa_z/2)\nonumber\\
&+c_3\,e^{\kappa_z}{\rm erfc}(\kappa_z),
\end{align}
where $K_n$ is the Macdonald function, ${\rm erfc}$ is the complementary error function, and
\begin{subequations}\label{coef}
\begin{align}
c_0&=\frac{204 - 175 \kappa_z + 120 \kappa_z^2 + 60 \kappa_z^3 + 12 \kappa_z^4}{12 \kappa_z^3},\\
c_1&=\frac{8 - 18 \kappa_z - 12 \kappa_z^2}{3 \kappa_z^{3/2}\sqrt{\pi}},\\
c_2&=\frac{-40 + 32 \kappa_z + 6 \kappa_z^2 + 12 \kappa_z^3}{3 \kappa_z^{5/2}\sqrt{\pi}},\\
c_3&=\sqrt{\pi}\frac{17 - 20 \kappa_z - 5 \kappa_z^2 - 24 \kappa_z^3 - 11 \kappa_z^4 - 2 \kappa_z^5}{2 \kappa_z^{7/2}}.
\end{align}
\end{subequations}
The results are shown in Fig.~\ref{fig4} where we plotted the exact function $I(\kappa_z)$ and its asymptotic form:
\begin{align}\label{asym}
I(\kappa_z)\approx \frac{33}{16\kappa_z^4},
\end{align}
obtained from the asymptotic expansions of the error function and the Macdonald functions (Eqs.~8.254 and 8.451 of \cite{gr}).


\begin{thebibliography}{00}
\bibitem{bb1} I. Bialynicki-Birula and Z. Bialynicka-Birula, Phys. Rev. Lett. {\bf 108}, 140401 (2012).
\bibitem{bb2} I. Bialynicki-Birula and Z. Bialynicka-Birula, Phys. Rev. A {\bf 86}, 022118 (2012).
\bibitem{sieg} A. E. Siegman, in {\it DPSS (Diode Pumped Solid State) Lasers: Applications and Issues}, M. Dowley, ed., Vol. 17 of OSA Trends in Optics and Photonics (Optical Society of America, 1998), paper MQ1.
\bibitem{al1} M. A. Alonso and G. W. Forbes, JOSA A {\bf 17}, 2391 (2000).
\bibitem{al2} M. A. Alonso, R. Borghi and M. Santarsiero, JOSA A {\bf 23}, 691 (2006).
\bibitem{al3} M. A. Alonso, R. Borghi and M. Santarsiero, JOSA A {\bf 23}, 701 (2006).
\bibitem{al4} M. A. Alonso J. Opt {\bf 13}, 064016 (2011).
\bibitem{bi} M. Born and L. Infeld, Proc. Roy. Soc. Lond. A {\bf 150}, 141 (1935).
\bibitem{js} J. Schwinger, Phys. Rev. {\bf 130}, 406 (1963).
\bibitem{rg} R. J. Glauber, Phys. Rev. {\bf 131}, 2766 (1963);
\bibitem{tg} U. M. Titulauer and R. J. Glauber, Phys. Rev. {\bf 145}, 1041 (1966);
\bibitem{sr} Brian J. Smith and M. G. Raymer, New J. Phys. {\bf 9}, 414 (2007).
\bibitem{pp} In our previous paper \cite{bb2} the symbol $\sintt$ did not include the factor $1/k$.
\bibitem{bb} I. Bialynicki-Birula and Z. Bialynicka-Birula, {\em Quantum Electrodynamics} (Pergamon, Oxford, 1975), Chap.~9;
\bibitem{bb3} I. Bialynicki-Birula and Z. Bialynicka-Birula, Phys. Rev. D {\bf 35}, 2383 (1987);
\bibitem{bb4} I. Bialynicki-Birula and Z. Bialynicka-Birula, J. Opt. {\bf 13}, 064014 (2011).
\bibitem{bb5} I. Bialynicki-Birula and Z. Bialynicka-Birula, J. Phys. A {\bf 46}, 053001 (2013).
\bibitem{bb6} I. Bialynicki-Birula and Z. Bialynicka-Birula, Opt. Comm. {\bf 264}, 342 (2006).
\bibitem{gr} I. S. Gradshteyn and I. M. Ryzhik, {\em Tables of Integrals, Series, and Products}, (Academic Press, New York, 2000).
\bibitem{math} Wolfram Research, Inc., Mathematica, Version 9, Champaign, IL (2012).
\bibitem{qdegl} S. Quabis, R. Dorn, M. Eberler, O. Gl¨ockl, and G. Leuchs, Opt.
Commun. {\bf 179}, 1 (2000);
\bibitem{yb} K. S. Youngworth and T. G. Brown, Opt. Express {\bf 7}, 77 (2000);
\bibitem{dql} R. Dorn, S. Quabis, and G. Leuchs, Phys. Rev. Lett. {\bf 91}, 233901
(2003);
\bibitem{yy} S. Yan and B. Yao, Phys. Rev. A {\bf 77}, 023827 (2008);
\bibitem{op} S. Orlov and U. Peschel, Phys. Rev, A {\bf 82}, 063820 (2010).
\end{thebibliography}
\end{document}